\documentclass[runningheads]{llncs}

\usepackage{graphicx}
\usepackage{booktabs} 

\begin{document}
\title{Lessons Learned from a Decade of Providing Interactive, On-Demand High Performance
Computing to Scientists and Engineers
\thanks{This material is based upon work supported by the Assistant Secretary of Defense for Research and Engineering under Air Force Contract No. FA8721-05-C-0002 and/or FA8702-15-D-0001. Any opinions, findings, conclusions or recommendations expressed in this material are those of the author(s) and do not necessarily reflect the views of the Assistant Secretary of Defense for Research and Engineering.}}
\titlerunning{Lessons Learned Providing Interactive HPC}

\author{Julia Mullen\inst{1} \and
Albert Reuther\inst{1} \and
William Arcand\inst{1} \and
Bill Bergeron\inst{1} \and
David Bestor\inst{1} \and
Chansup Byun\inst{1} \and
Vijay Gadepally\inst{1,2} \and
Michael Houle\inst{1} \and
Matthew Hubbell\inst{1} \and
Michael Jones\inst{1} \and
Anna Klein\inst{1} \and
Peter Michaleas\inst{1} \and
Lauren Milechin\inst{2} \and
Andrew Prout\inst{1} \and
Antonio Rosa\inst{1} \and
Siddharth Samsi\inst{1} \and
Charles Yee\inst{1} \and
Jeremy Kepner\inst{1,2}}


\authorrunning{J. Mullen, A. Reuther, et al.}
%
\institute{MIT Lincoln Laboratory, 244 Wood Street, Lexington, Massachusetts 02420, USA 
\email{\{jsm, reuther, warchand, bbergeron, david.bestor, byun, vijayg, michael.houle, anna.klein, pmichaleas, aprout, antonio.rosa, sid, yee, kepner\}@LL.mit.edu}\\
\and
Massachusetts Institute of Technology, 77 Massachusetts Avenue, Cambridge, Massachusetts 02139, USA\\
\email{lauren.milechin@mit.edu}}
\maketitle              

\begin{abstract}
For decades, the use of HPC systems was limited to those in the physical sciences who had mastered their domain in conjunction with a deep understanding of HPC architectures and algorithms.  During these same decades, consumer computing device advances produced tablets and smartphones that allow millions of children to interactively develop and share code projects across the globe.  As the HPC community faces the challenges associated with guiding researchers from disciplines using high productivity interactive tools to effective use of HPC systems, it seems appropriate to revisit the assumptions surrounding the necessary skills required for access to large computational systems.  For over a decade, MIT Lincoln Laboratory has been supporting interactive, on-demand high performance computing by seamlessly integrating familiar high productivity tools to provide users with an increased number of design turns, rapid prototyping capability, and faster time to insight.   In this paper, we discuss the lessons learned while supporting interactive, on-demand high performance computing from the perspectives of the users and the team supporting the users and the system.   Building on these lessons, we present an overview of current needs and the technical solutions we are building to lower the barrier to entry for new users from the humanities, social, and biological sciences.

\end{abstract}

\section{Introduction}
\label{Introduction}

Traditionally supercomputers and high performance computing (HPC) were the domain of experts with deep understanding of their scientific discipline, computer architecture and software.  For virtually the entire history of HPC, the standard path to developing the skills necessary for supercomputer usage included graduate programs in the physical sciences and engineering.  These research programs, with applications requiring massive computational effort, prepared students who had the time, inclination and mandate from their advisors to learn how to program and exploit the computational power of supercomputers.  Many of these graduates went on to research positions at centers, research laboratories, and universities where they trained the next generation of HPC researchers thereby growing the HPC community and reinforcing the notion of a single path to HPC.

This single learning path may have been appropriate when computational approaches to science, engineering and design were in their infancy, but today 
computational approaches to problem solving have become commonplace, used by researchers from engineering and the physical sciences and more recently by members of the medical and social sciences and the humanities.  While many engineering and  science disciplines have long included computing in their undergraduate and graduate programs, such academic preparation is minimal to non-existent for many of the disciplines that have recently adopted computational solution strategies.  For these research communities the single path to HPC is a significant deterrent.  Furthermore, industry relies on the use of HPC systems to stay globally competitive~\cite{CTWatch} but the majority of the industrial workforce has little or no HPC experience.  Developing HPC experience among this portion of the workforce calls for new learning paths designed for mid-career professionals.
 

This friction between the steep requirements of a single pathway to HPC use and the need for high productivity HPC systems is not new.  As few as 15 years ago, prior to the development of MathWorks' Parallel MATLAB\textsuperscript{\textregistered} product and the parallel versions of standard engineering design tools, {\em e.g.}, Fluent, Ansys, NASTRAN, etc., the science and engineering community faced the same divide: a small number of researchers became expert HPC users while the vast majority of scientists and engineers used interactive, high productivity tools on their desktop systems, upgrading hardware when greater performance was required or available.  However, as processor clock rates stagnated and hardware improvements because modestly incremental, hardware upgrades no longer delivered large gains. From this experience of the breaking down of Moore's Law, many have begun to realize that HPC is required.

It is easy to understand the limits of this single learning path for the broad community of scientists and engineers who faced increasingly complicated applications but no clear path to merging the productivity of interactive tools with the performance of compute clusters and supercomputers.  For these researchers the traditional path to greater performance required building the software frameworks in a compiled language, learning to create batch scripts, and becoming accustomed to the software testing and development delays associated with batch systems.  For many, the costs associated with these changes in terms of both time and distraction were perceived to be greater than the reward.  For them, another approach was necessary.  

At MIT Lincoln Laboratory a standard prototyping process for analysts includes developing, testing and debugging in MATLAB\textsuperscript{\textregistered} to speed up the design and prototyping phase before passing the engineering code to a team of expert real time coders.  The real time team is responsible for converting the MATLAB\textsuperscript{\textregistered} to C or $C^{++}$ code and tuning it for the target architecture~\cite{VSIPL}. This division of labor is ideal for certain situations, such as very large, well funded projects,  but is not generally feasible for smaller companies and teams.  For smaller, time critical projects, a merger  of high productivity and high performance is essential.

For almost 15 years the Lincoln Laboratory Supercomputing Center (LLSC, formerly the Lincoln Laboratory Grid (LLGrid) team) at MIT Lincoln Laboratory has provided interactive on-demand cluster computing resources to over 1,000 researchers at the Laboratory~\cite{LLJournal}. As part of the LLSC mission to deliver new and innovative technologies and methods, we have developed and built the MIT SuperCloud~\cite{reuther2013llsupercloud} to enabling scientists and engineers to quickly ramp up the pace of their research and rapid prototyping by leveraging big compute and big data storage assets. The SuperCloud is a fusion of the four large computing ecosystems: supercomputing, enterprise computing, big data and traditional databases into a coherent, unified platform. The MIT SuperCloud has spurred the development of a number of cross-ecosystem innovations in high performance databases~\cite{byun2012driving}, \cite{kepner2014achieving}; database management~\cite{prout2015enabling}; data protection~\cite{kepner2014computing}; database federation~\cite{kepner2013d4m}, \cite{gadepally2015d4m}; data analytics~\cite{kepner2012dynamic}; dynamic virtual machines~\cite{reuther2012hpc-vms}, \cite{jones2016scalability} and system monitoring~\cite{hubbell2015big}.

In general, interactive, on-demand supercomputing is very useful for a variety of research, engineering, and prototyping activities including algorithm development, data analysis, machine learning training, application steering, and visualization. Over the past 15 years, the common Laboratory use cases encompass many of these activities and have included algorithm development for sensor signal processing; development of multiple program, multiple data (MPMD) real time signal processing systems; high throughput computing for aircraft collision avoidance system testing; biomedical analytics to develop medical support techniques for personnel in remote areas; and prototyping capabilities for a range of systems.  Unlike traditional HPC applications, most of these capabilities involve prototyping efforts for multi-year, but not multi-decade, mission-driven programs making it even more important that researchers are able to use familiar interactive tools and achieve a greater number of design cycles per day. To enable an interactive high performance development environment, our team turned the traditional HPC paradigm on its head.  Rather than providing a batch system, training in MPI, and assistance porting serial code to a supercomputer, we developed the tools and training to bring HPC capabilities to the researchers' desktops and laptops.  As common use cases and staff computational preparation change, we routinely update our tools so that we can provide relevant interactive research computing environments.  From our initial experience, standing up an interactive on-demand cluster computing resource through our current support of machine learning, data analytics and user portals, we have focused on creating multiple paths to HPC usage.  In this paper, we present and discuss the lessons we have learned and how they apply to the larger HPC ecosystem.  

The paper is organized as follows, we present the high-level lessons that we have learned in Section~\ref{sec:lessonIntro}.  In Section~\ref{sec:approach} we dive deeper into the lessons we have learned about provisioning an appropriate system, creating a software abstraction layer and providing the training required to support interactive HPC.  Section~\ref{sec:metrics} considers standard HPC metrics and the reframing necessary to create metrics that capture the value of interactive HPC for smaller centers, universities, and industry.  We close with a summary.

\section{Lessons Learned}
\label{sec:lessonIntro}

The key elements required to provide interactive on-demand HPC to a user base spanning neophytes to experts can be summed by four high level ideas. These four key elements are:  broadening the definition of interactive within the HPC community; expanding the HPC ecosystem; re-architect the HPC system, where by system we include system architecture, software stack, and user support; and reframing the success metrics.  Two of these key elements, the idea related to broadening the definition of interactive within the HPC community and the related idea of expanding the HPC community are philosophical in nature and are described within this section. The remaining elements are primarily structural and require a fuller description.  These elements, the manner in which we re-architect the HPC system and reframing the success metric are described in this section and detailed in Sections~\ref{sec:approach} and \ref{sec:metrics}, respectively.

\subsection{Broadening the Definition of Interactive HPC}
First and foremost, we recognize that there is a large middle ground of users who want computational environments that balance performance and usability.  In practice this translates to redefining the term ``interactive High Performance Computing''.  Virtually every center sets aside a partition of  compute cores for debugging and interactive use during normal working hours.  Generally,  these interactive partitions provide access to a command shell where a user can build, submit and track a compute job without having to interface through a batch scheduler or specifically request resources.   For an expert HPC user familiar with the Linux command line and batch processing workflows, this partition provides a reasonable way to  debug compiled code and scripts, including submission scripts, on a small scale prior to launching production level jobs.   However, for the researcher who is familiar with a modern Integrated Development Environment (IDE), such a workflow is convoluted, confusing and opaque.  A first step in supporting the work of a more general researcher base is to recognize the extensive use and value of IDEs and develop ways to bring them into the HPC environment.  When the LLGrid project started, there were no widespread software tools that connected the IDE at a desktop to more robust compute resources. Over the past decade or so, pMatlab~\cite{LLJournal}, StarP~\cite{starp-edelman} and MATLAB{\textregistered} Distributed Compute Server (MDCS) have filled this void for researchers using MATLAB{\textregistered} while other commercial software products have created versions of their products that run seamlessly on parallel or high throughput systems but present the user with a familiar front end  (\emph{e.g.}, Julia, Python, Mathematica, Fluent, NASTRAN, Anysys, and FEKO). For these applications and users, interactive assumes that there is a desktop or browser user interface, and the user will be able to simply hit the return key and the job will launch and run immediately.   This is interactive and on-demand, matching the interactive desktop experience but with greater memory, compute and network resources.  Starting from this broader definition of interactive HPC it is much easier to design the appropriate architecture and develop the necessary middleware tools to expand the HPC user community.

\subsection{Re-architecting for Interactive HPC}
At its essence, creating an interactive on-demand HPC environment means bridging the gap between standard HPC architectures and the user's desktop experience.  An interactive on-demand system, like any HPC system is built with login nodes, a scheduler, compute nodes, a shared central file system and a network.  The particular hardware selection and configuration --- i.e., the amount of RAM, number of cores, network technology and topology, etc. --- is a tradeoff between cost and the requirements of a set of common user applications.  
While the testing confirms that the applications will run, providing a compute system is not enough to attract users from beyond the expert user base trained through the graduate student pathway.  To increase adoption from a broader segment of the scientific, engineering, business, social science and humanities domains as well as the mid-career professionals in industry we need to provide an HPC framework that is approachable and reliable.   The components of such a framework include the compute system mentioned above along with an OS and the necessary systems tools, which we will call the system, and a layer of software, or middleware, between the system and user applications, which we will call the software.  This middleware layer is where the technical challenges of supporting broader computing communities arise and where HPC expertise is essential to creating tools to lower the barrier to entry. Section~\ref{sec:approach} details the effective solution we have developed and particular lessons we have learned through their deployment.

\subsection{Reframing the Metrics of Success}
\label{sec:reframing}
Virtually all HPC centers report the percentage of system utilization as their metric of success.  This choice of metrics often leads to queuing systems and user behavior designed to feed the system with the type of jobs that yield high utilization.  However, these utilization-based queuing practices are often at odds with rapid prototyping of algorithms and simulations, exploration of large datasets and real time steering of complicated multi-physics simulations~\cite{reuther2007technical}. Job queuing systems are configured to accumulate and maintain a backlog of large and small jobs so that as soon as an executing job completes, one or more jobs can replace it and execute. This encourages user behavior that includes submitting a multitude of jobs with many different parameters thus accumulating even more jobs in the queue backlog.  Section~\ref{sec:metrics} revisits work for DARPA's High Productivity Computing System program where a productivity metric was developed as part of a larger analysis of HPC Return On Investment for a broad range of applications and research domains.

\subsection{Expanding the HPC Ecosystem}

\begin{figure}
  \includegraphics[width=5in]{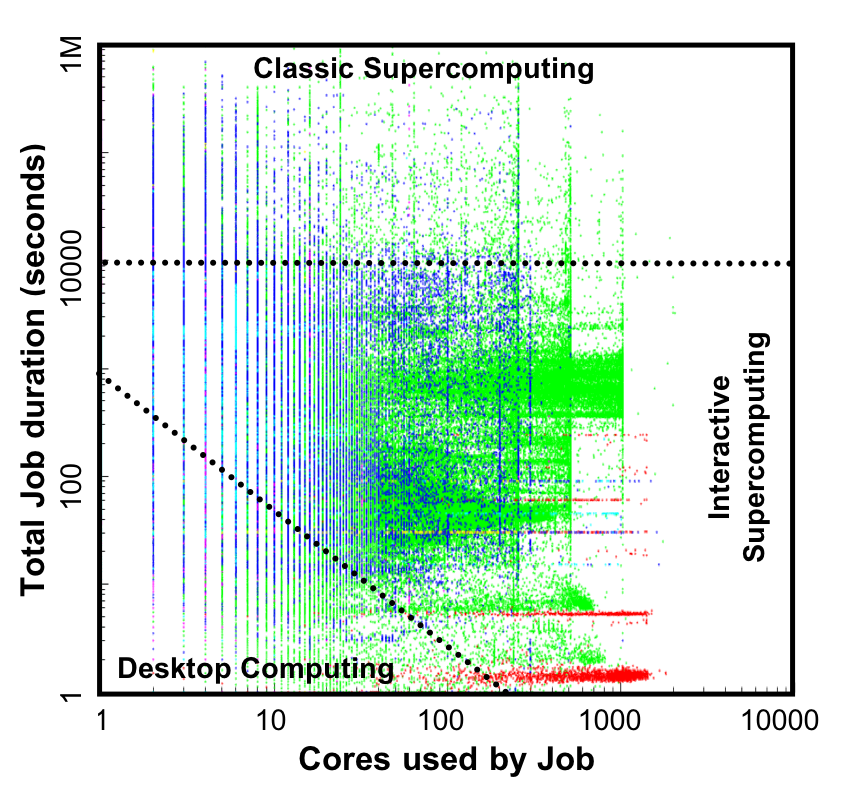}
  \caption{A notional view of research computing regimes. The Desktop Computing region involves jobs that execute in less than five minutes, often for exploration and debugging. The Classic Supercomputing jobs are those that run for over three hours. Finally, the Interactive Supercomputing jobs run for between five minutes and three hours and are usually jobs that involve rapid turnarounds for insight.}
  \label{fig:regimes}
\end{figure}

There will always be a place for large, batch processing systems that provide expert users the resources to attain the best performance for a given application or analytic.  The case that we make here is not that all systems need to be interactive, but rather that we need to expand our vision of what an HPC ecosystem includes and how it supports research, design and prototyping at all levels.  Across a spectrum of applications and centers, the weight given to performance versus productivity will and should vary based on the user applications to incorporate the computing regimes illustrated in Figure~\ref{fig:regimes}.  This approach means that as we reach out to new communities familiar with IDEs and workflows that incorporate research portals we need to rethink our presumed prerequisites.  The community has put significant effort into teaching new users the basics of Linux and the command line interface and the gritty details of batch scheduler systems and MPI~\cite{swcarpentry}, \cite{wilson2006software} --- effort that reinforces the belief that new user communities must adopt our workflows and tools.  These efforts are attempts at building shortcuts or mid-career bootcamps toward our single educational path to HPC.  Re-evaluating our assumptions of what an HPC ecosystem includes offers the opportunity for HPC experts to apply Design Thinking~\cite{designThinkingBootcamp} to the design of new paths to HPC usage for communities who require not just performance but a balance of productivity and performance.   The first step in Design Thinking is to capture and understand the user perspective, goals and workflow; starting with this step increases the likelihood that tools and training will lead to additional pathways to HPC. 

\section{Architecture Requirements for Interactive HPC}
\label{sec:approach}

As the community begins capturing the needs of these communities with an eye toward expanding the number of pathways to HPC resources, 
it is important to remember how we addressed a similar challenge:  bringing an analyst and research community to using high productivity languages and interactive tools into the parallel and distributed computing environment.  In the early 2000s, MIT Lincoln  Laboratory staff were seeing  steady increases in the fidelity and capability of sensors, leading to increasingly sophisticated sensor signal and image processing algorithms~\cite{LLJournal}.  The level of complexity, combined with the sharp increase in data to be processed and the end of Moore's Law~\cite{Moore} meant that desktop workstations were no longer able to provide adequate computational power.  To advance their work, researchers and analysts had access to compute clusters, but these generally required relinquishing the interactive environment and converting application software written in high productivity languages to compiled languages.  For many,  the performance gain associated with compiled parallel code did not override the ease of use and rapid prototyping capabilities of interactive languages, especially for mission driven projects focused on design and algorithm prototypes.  

In 2003, as grid computing was emerging, the question was, can systems carefully configured for high productivity and rapid prototyping  fill the gap between slower desktops and big HPC systems and support the growing needs of analysts and researchers for whom the traditional HPC learning path was a barrier to entry?  To evaluate this question,  the Lincoln Laboratory Grid (LLGrid, now LLSC) was created to explore and develop interactive, on-demand high performance computing for the Laboratory.

\subsection{System}
Bringing a systems engineering approach to the challenge of designing an interactive, high productivity, on-demand high performance computing environment, we began by identifying the three subsystems that form the environment: the compute cluster; the software stack (particularly the middleware layer); and the development of user consultation and support unit.  In terms of hardware selection and design, our HPC system was similar to most small to moderately-sized systems. In 2003-2005, we built our first clusters using commodity-off-the-shelf (COTS) system components including dual-socket, single-core compute nodes, gigabit Ethernet interconnects, a shared central RAID file system exported with the NFS and CIFS protocol standards, and the University of Wisconsin Condor scheduler.  As is always the case, these components were chosen by balancing the cost against user application requirements. What differentiated our HPC system from others was not the choice of components, but the configuration of the scheduler.  To provide on-demand computing services meant rethinking the traditional partition configurations: each user was limited to running jobs up to a core-count limit equaling approximately one eighth of the total cores available. They could request that their core allocation limit be increased for a finite time period via email to our administration team. This limit usually assured that a subset of the system's compute cores were always available. Also the scheduler and central file system were configured and tuned so that launches of parallel jobs occurred in less than 20 seconds on hundreds of cores, thereby providing the interactivity with job launches that users were used to on their desktop IDEs~\cite{reuther2007technical}. 

\begin{figure}
  \includegraphics[width=5in]{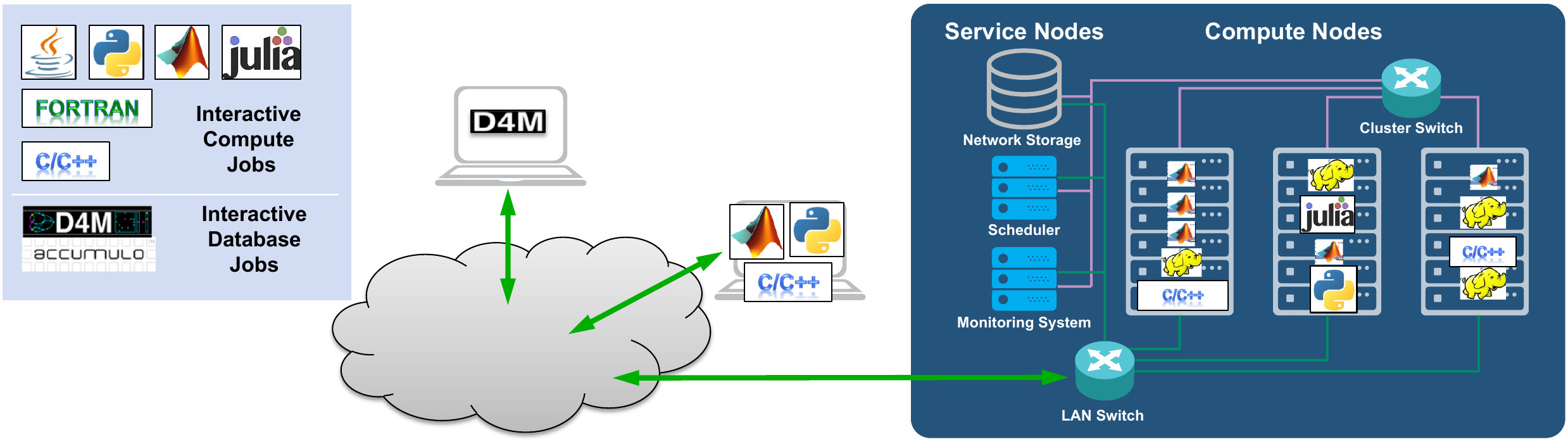}
  \caption{The unified SuperCloud architecture and compute platform.}  \label{fig:unified}
\end{figure}

In subsequent clusters that we have built, we have used similar components as technology has progressed eventually including multicore (up to 64 core) compute nodes. Our latest systems have 10-gigabit Ethernet connections with a 1024-port non-blocking central core switch, along with a gigabit out-of-band management network. And the central file system has thousands of disk drives, many levels of redundancy, and can read and write tens of gigabytes of data per second. But hardware has not been the only place for improvements. 
Over the years, sensor and processing capabilities continued to increase, driving interest in using the compute system for data analysis applications.  To accommodate the requirements (memory, storage, tools) of these applications, our traditional HPC system was extended to include High Performance Databases such as Accumulo and SciDB~\cite{byun2012driving}. The databases are dynamically run on a set of compute nodes, creating a unified compute platform for the users as depicted in Figure\ref{fig:unified}.   Researcher access to start, stop and monitor these databases was integrated into the system via a web-based portal, and software tools were created to easily connect databases and user applications.   As the user application space expanded to include machine learning and medical science applications from users without traditional computing experience, the unified platform was extended to include portal interfaces~\cite{portal} to support web applications, particularly ones that integrate into sensor processing and scientific computing workflows. As we extend the unified platform into these new capabilities, we will further develop and adapt all of our technical and support efforts to effectively enable users. 


\subsection{Software}
Providing hardware is not enough, particularly for researchers whose academic training did not include HPC topics or exposure.  Engaging new HPC users requires filling the gap between interactive high productivity programming and the high performance compute hardware.  This middleware layer is crucial to the success of 
mainstream interactive HPC. In 2003, when roughly 85$\%$ of computational engineering and science was accomplished using MATLAB\textsuperscript{\textregistered}, there was not only no communication library for the language but there was no plan to create a communication library~\cite{MolerDiscussion}. If research analysts using MATLAB\textsuperscript{\textregistered} were going to take advantage of a distributed cluster, they needed middleware to enable the parallel capability.  As a first step, Dr. Jeremy Kepner created MatlabMPI,~\cite{matlabMPI} a communication layer  that used the file system as a communication fabric and leveraged the save and load functions in MATLAB\textsuperscript{\textregistered} to execute  send and receive commands.  The choice of communication fabric was a design decision to retain MATLAB\textsuperscript{\textregistered}'s platform agnostic feature and support the range of desktop systems in use at MIT Lincoln Laboratory and  the broader science and engineering community.  While MatlabMPI enabled parallel MATLAB applications, it had the same drawback as MPI, in that the programmer needed to spend significant effort to manage the communication.  To separate the parallel programming details from the application programming concerns, Lincoln Laboratory created pMatlab~\cite{pmatlab}, \cite{pmatlabbook}.  pMatlab implemented Partitioned Global Address Spaces (PGAS) constructs so that the research programmer could design the application on the global level but use pMatlab constructs to distribute the data structures, manage the local-global mapping and communication.  

\begin{figure}
  \includegraphics[width=5in]{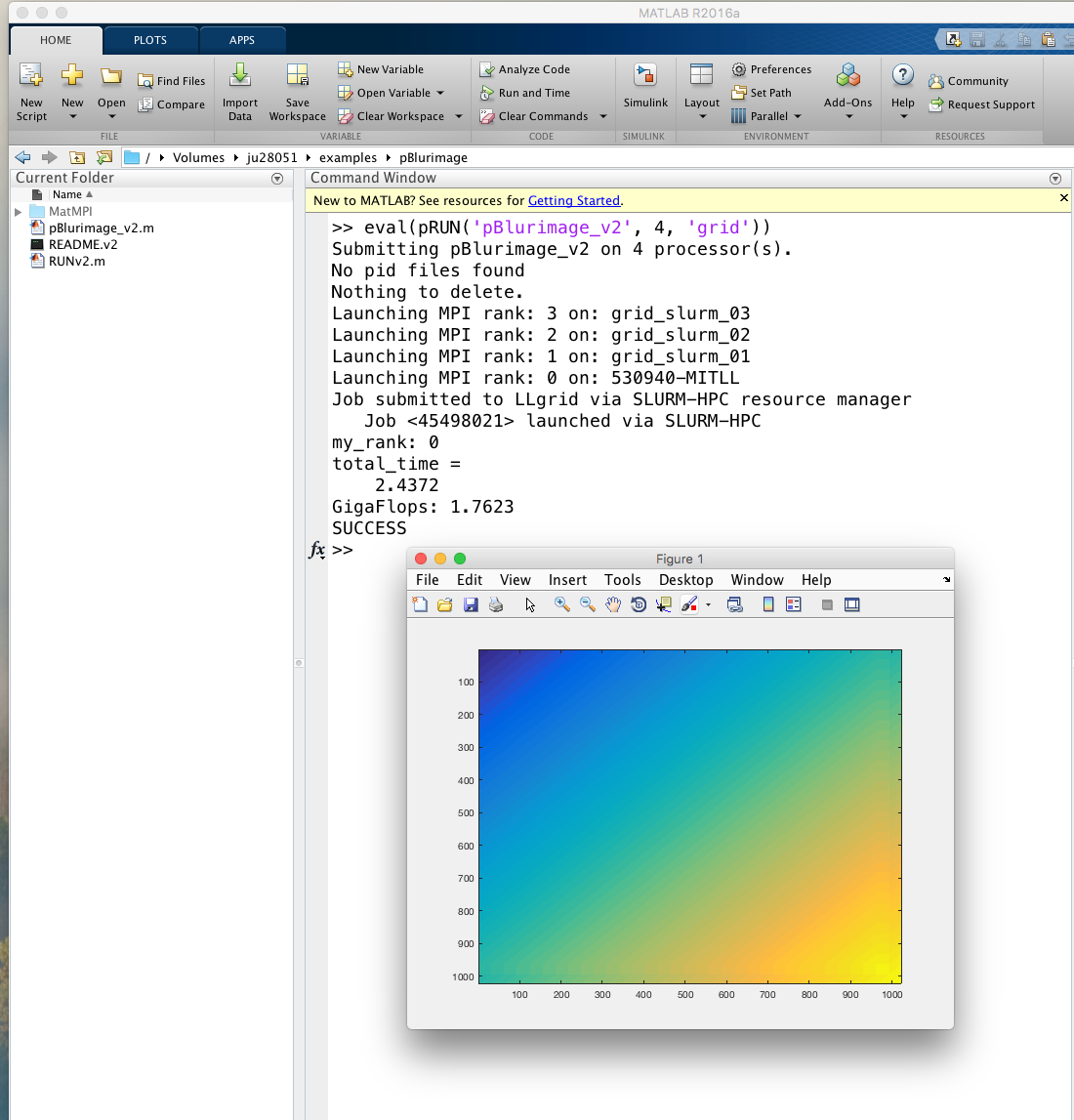}
  \caption{Running a pMatlab job on four processors.  The resulting image is gathered to Node 0, the desktop, for display.}  \label{fig:matlab-pBlur}
\end{figure}

The pMatlab library provided a high productivity ease-of-use approach to writing distributed computing applications, but it did not address the challenges of batch schedulers and job submission on shared cluster resources.  The scheduler and job submission challenge was addressed by Dr. Albert Reuther through gridMatlab~\cite{gridMatlab}, a set of tools, written in MATLAB\textsuperscript{\textregistered}, that provides the glue between the researcher's desktop and the compute cluster.   Researchers submit their compute jobs using an overloaded version of the MATLAB\textsuperscript{\textregistered} \texttt{eval} command that accepts three arguments: the name of the script to be run, the number of cores and the location of the cores, {\em i.e.} the local desktop; use of the local desktop along with the cluster; or as a background job on the cluster.  The default is to assign node 0 to the researcher's desktop with all other nodes on the cluster.  One benefit of this setting is that results are aggregated to node 0 local desktop and can easily be post-processed on the researcher desktop.  This provides the user with a familiar interface for job launches versus traditional command line tools.   Figure~\ref{fig:matlab-pBlur} illustrates the launch of a simple example on four processors.  The reserved word 'grid' indicates that the job will be run with 3 processors on the cluster and one on the local desktop, as seen in the echo of each launch command.  The job is submitted to the scheduler and when complete the results are written to the local MATLAB\textsuperscript{\textregistered} command window.  At no time in the process does the researcher log into the cluster; all of the development work and job submission is done from the Integrated Development Environment (IDE) on the user's desktop or laptop.  When the job completes, or fails, the researcher has immediate feedback and can quickly modify the application and resubmit, offering a seamless prototyping environment.

This approach to developing middleware is key to HPC adoption by a broader range of research disciplines and should be a focus of the HPC experts.  The philosophy behind pMatlab and gridMatlab is that HPC experts can create tools to abstract the tricky parallel details that can sidetrack beginner or intermediate HPC programmer progress thus lowering the barrier to entry for new users.  In keeping with this philosophy, new tools for abstraction and ease of use were created to support new data analysis and machine learning applications.

\subsection{Supporting Users}
Building an effective tool does not guarantee adoption and what seems easy to the expert many not be intuitive to a novice or intermediate.  To insure adoption, the LLSC team took a novel approach to on-boarding new HPC users by providing targeted individual tutorials to each new user.  The tutorial included a general introduction to high performance and parallel computing,  a careful walk through the process of building a parallel parameter sweep application from a serial code and a discussion of parallelization strategies for the user application.  Over the years we have gathered much of this educational content and created a MOOC course that runs on the HPC system and provides the user with course examples to run on the HPC system thus building understanding of both general HPC concepts and the specifics of applications on our system~\cite{mullen2017learning}. 

More recently we have begun developing our examples within Jupyter notebooks because they provide a method to create sections that explain the concepts and tasks interleaved with sections where students can edit and run code.  Consistent with our other tools and services ({\em e.g.}, pMatlab and databases), the Jupyter notebook compute engines are run on compute nodes of our HPC system so that the educational examples are running directly on the system allowing users the opportunity to test and explore HPC strategies. All the while, the graphical interface of the Jupyter notebooks are executing in the browser of the user's desktop or laptop computer~\cite{portal}.

\section{Metrics}
\label{sec:metrics}

After covering all of these system and user support topics, we now return to discussing metrics. Using the proper metrics is very valuable for determining whether an organization is getting the most out of their HPC investments. This leads to better leadership understanding of the value of HPC for the organization and encourages each organization to manage towards those goals that most benefit the organization.  

As we briefly discussed in Section~\ref{sec:reframing}, managing toward maximum system utilization is directly at odds with enabling interactive, on-demand HPC. The DARPA High Productivity Computing Systems (HPCS) program brought this discrepancy to full light, and it was on through this program that a new method to measure the return on investment (ROI) was developed~\cite{CTWatch}. The ROI calculation places the benefit in the numerator and sums all of the costs in the denominator. Each organization can determine what the benefits for their use cases are, and they can compile all of the costs for the denominator. Examples of benefits at various organizations are presented in \cite{CTWatch}. At Lincoln Laboratory, we chose to measure the benefit as the sum of all the time saved by users on a system by running parallel jobs over running single-process jobs. We enumerated the cost of enabling HPC at Lincoln to be the sum of the time to parallelize each code set, the time to train users, the time to launch jobs, the time to administrate the system, and the system cost. Details of our analysis can be found in~\cite{CTWatch} and~\cite{reuther2007technical}. We were pleased to find that the time saved often help us achieve organizational ROI of 2x to 10x (or more) -- clearly, a very beneficial values for our Laboratory. Some of this can be attributed to very efficient system administration workflows and low launching and development overhead. However, a large part of the benefit is due the broad utilization across the entire Laboratory and the value it has brought to each of the users.

\section{Summary and Future Work}

In summary, the challenges to deploying interactive on-demand HPC environments are both technical and institutional. The technical challenges involve developing middleware at the correct level of abstraction to lower the slope of the learning curve for new users and provide a path to increased productivity. The human challenges center on the development of a community of practice that appreciates the importance of balancing performance and productivity, a re-evaluation of the assumptions surrounding the metrics of success and the creation of educational materials aimed at building new pathways to HPC expertise. Together these changes advance a new approach for provisioning HPC environments.
The strategies presented here are general and easily adapted for any center where productivity and the rapid prototyping and testing of algorithms and analytics are key concerns. The challenge going forward is to recognize emerging needs from both new users and new domains and create the appropriate middleware and educational materials. While we started by abstracting the parallel and scheduling details of launching parallel MATLAB\textsuperscript{\textregistered} in the early 2000’s, we have since added HPC databases for analytics and Jupyter notebooks to abstract the command line concerns and launch details for new languages such as Julia, Python and R. As the user community and their applications change, we will continue to evaluate their preferred tools, languages and environments against the demands of the HPC system to uncover areas where abstractions, system efficiencies and education can yield both productivity and performance.

\bibliographystyle{splncs04}
\bibliography{MullenWIHPC18}

\end{document}